\documentclass[10pt,a4paper]{article}

\usepackage[margin=25mm]{geometry}
\usepackage{graphicx}
\usepackage{csquotes}
\usepackage[english]{babel}
\usepackage[T1]{fontenc}
\usepackage[utf8]{inputenc}
\usepackage[all]{nowidow}
\usepackage{xurl}
\usepackage{subcaption}
\usepackage{tabularray}
\UseTblrLibrary{booktabs}
\usepackage{tikz}
\usetikzlibrary {arrows.meta}
\usepackage{fancyvrb} 
\usepackage{times}
\usepackage{authblk}

\usepackage{float}
\usepackage[colorlinks=true, allcolors=darkgray]{hyperref}

\usepackage[%
	maxbibnames=999,          
	maxcitenames=999,         
	isbn=false,               
	backend=biber]{biblatex-chicago}
\usepackage{fvextra} 
\usepackage[export]{adjustbox} 
\newcommand{\surligne}[1]{{\setlength{\fboxsep}{0pt}\colorbox{yellow}{\strut #1}}}

\title{From Show Programs to Linked Data. A Generative AI Workflow for Performing Arts Collections\thanks{\fullcite{bardiotShowProgramsLinked2026}}}

\author[1]{Clarisse Bardiot}
\author[2]{Bernard Jacquemin}
\author[1]{Jacob Hart}
\author[3]{George Bruseker}
\author[1]{Antonios Lagarias}
\author[4]{Pierre-Carl Langlais}
\author[1]{Brandon Farnsworth}
\author[1]{Jeanne Fras}

\affil[1]{Univ Rennes, APP (Arts: Pratiques et Poétiques), F-35000 Rennes, France --- {first.last}@univ-rennes2.fr}
\affil[2]{Enssib, EA 4147 Élico, F-69100 Villeurbanne, France --- Bernard.Jacquemin@enssib.fr}
\affil[3]{Takin Solutions / Swiss Art Research Infrastructure --- george@takin.solutions}
\affil[4]{Pleias, Associate researcher Medialab/SCAI, Paris, France --- pierrecarl.langlais@gmail.com}

\date{\vspace{-4ex}}

\begin{document}

\maketitle

\begin{abstract}
	\noindent Many heritage institutions hold extensive collections of show programs. Too numerous to be cataloged individually, they remain difficult to access and are still largely undigitized. This paper presents a workflow for transforming such documents into structured, interoperable data by combining vision-language models (VLMs), a custom extension of the Linked Art ontology for the performing arts, and different approaches for automatic semantic data annotation. Using the Festival d’Avignon programs (1947–present), preserved at the Bibliothèque nationale de France, as a case study, we demonstrate how VLMs achieve more than 98\% text transcription accuracy from heterogeneous heritage document images and how constraining large language model generation through a knowledge graph significantly reduces hallucinations in semantic annotation. We argue that this workflow opens new possibilities for large-scale, interoperable performing arts historiography, including the creation of \emph{catalogues raisonnés} for performing artists, a form of scholarship common in visual arts but absent in performing arts studies. 
\end{abstract}
%
\section{Introduction}

Show programs constitute one of the most crucial yet underused sources for performing arts historians. In the absence of other documentation, they frequently serve as the sole evidence that a performance occurred. Distributed to spectators on the day of the performance and sometimes hastily printed, these materials were not intended for preservation. Yet these ephemera, collected across numerous heritage institutions,\footfullcite{berbainPreservingTheaterEphemera2014}
record invaluable data about performances (cast and crew details, job titles, venue information, and last-minute modifications) while simultaneously revealing artistic networks, editorial strategies, and institutional dynamics. Despite this documentary richness, these documents remain largely inaccessible, as their sheer volume makes systematic cataloging and digitization impractical at scale.

Given that theater programs clearly define categories such as a show's title, author, director, cast, and venue, they may appear to be essentially structured data already. However, these categories often pose significant definitional challenges and risk anachronisms, particularly regarding performing arts professions, starting with the very notion of ``director.''\footfullcite{bardiotPerformingArtsDigital2021}
Moreover, performances represent complex artistic objects involving numerous individuals in diverse roles, programmed in different venues and on different dates and generating various forms of documentation, such as written publications, photographs, or films. They may also go through different states, such as ``work in progress.'' These complexities have impeded the development of unified ontological models for performance description, complicating interoperability among existing databases.\footfullcite{roussillonDossierEcrireLhistoire2020, vanoortStructuredDataPerforming2020, bardiotPerformingArtsDigital2021, douguetDossierBaseCESAR2024}

When programs are included in existing manually cataloged databases, they are typically reduced to minimal metadata, failing to capture the contextual information embedded in their full text. Manual processing of large program corpora at scale is prohibitively time-consuming for libraries, archives, and museums. Automating the task is equally challenging: standard optical character recognition (OCR) systems perform poorly on the complex, multilingual, and multicolumn layouts typical of programs; and converting unstructured content into standardized, machine-readable data requires ontological modeling expertise that automated tools have, until recently, handled unreliably. This bottleneck has a direct epistemological consequence: without access to full, structured program data, large-scale comparative research in performing arts history remains out of reach.

Recent advances in multimodal large language models (LLMs) and in ontology-driven reasoning offer new possibilities for addressing this challenge. This paper presents a concrete, reproducible workflow developed within the European Research Council (ERC) Advanced Grant project STAGE (From Stage to Data, 2024--2028) and based on a case study: the show program corpus of the Festival d'Avignon (1947--present), preserved at the Bibliothèque nationale de France (BnF). With more than 2,000 performances staged by international companies since its inception, the Festival d'Avignon has featured essential artists of the post-WWII era across theater, dance, performance, and circus. It serves as an ideal laboratory for examining the evolution of contemporary performing arts practices and networks. Our corpus, almost exhaustive, comprises approximately 1,840 show programs, covering 68 years (1954--2022) across two distinct periods of the Festival d'Avignon's history.\footnote{The corpus currently excludes years prior to 1954, as the festival's show program format was only introduced in that year. The scope of the study is also limited to the most recent completed directorship, which ended in 2022 under Olivier Py. In the short term, we plan to expand the corpus by incorporating season programs, which compile all events for each edition (including auxiliary activities such as concerts), adding approximately a bit more than 300 additional documents.}

The workflow combines three components: (1) a vision-language model (VLM)-based transcription outperforming traditional OCR, (2) a domain ontology extending Linked Art for the performing arts, and (3) different approaches to generate automatic semantic annotations constrained by this ontology. Thanks to this pipeline, it is possible to both faithfully transcribe heterogeneous historical documents and to parse their content into structured, machine-readable datasets. This approach enables new analytical possibilities: tracing how artistic communities form and evolve, mapping the role of cultural institutions in shaping contemporary performance practices, and supporting comparative studies across periods and geographical contexts. Beyond a single case study, we advocate for the collaborative construction of a standardized corpus of show programs, which would unlock numerous research avenues in performing arts history. Questions regarding the long-term formation of artistic networks, the circulation of companies across venues and national borders, or the construction of individual careers cannot yet be addressed systematically, precisely because program data have never been made available in structured, interoperable, machine-readable form at sufficient scale.

\begin{figure}[htbp]
	\centering
	\includegraphics[width=0.95\linewidth]{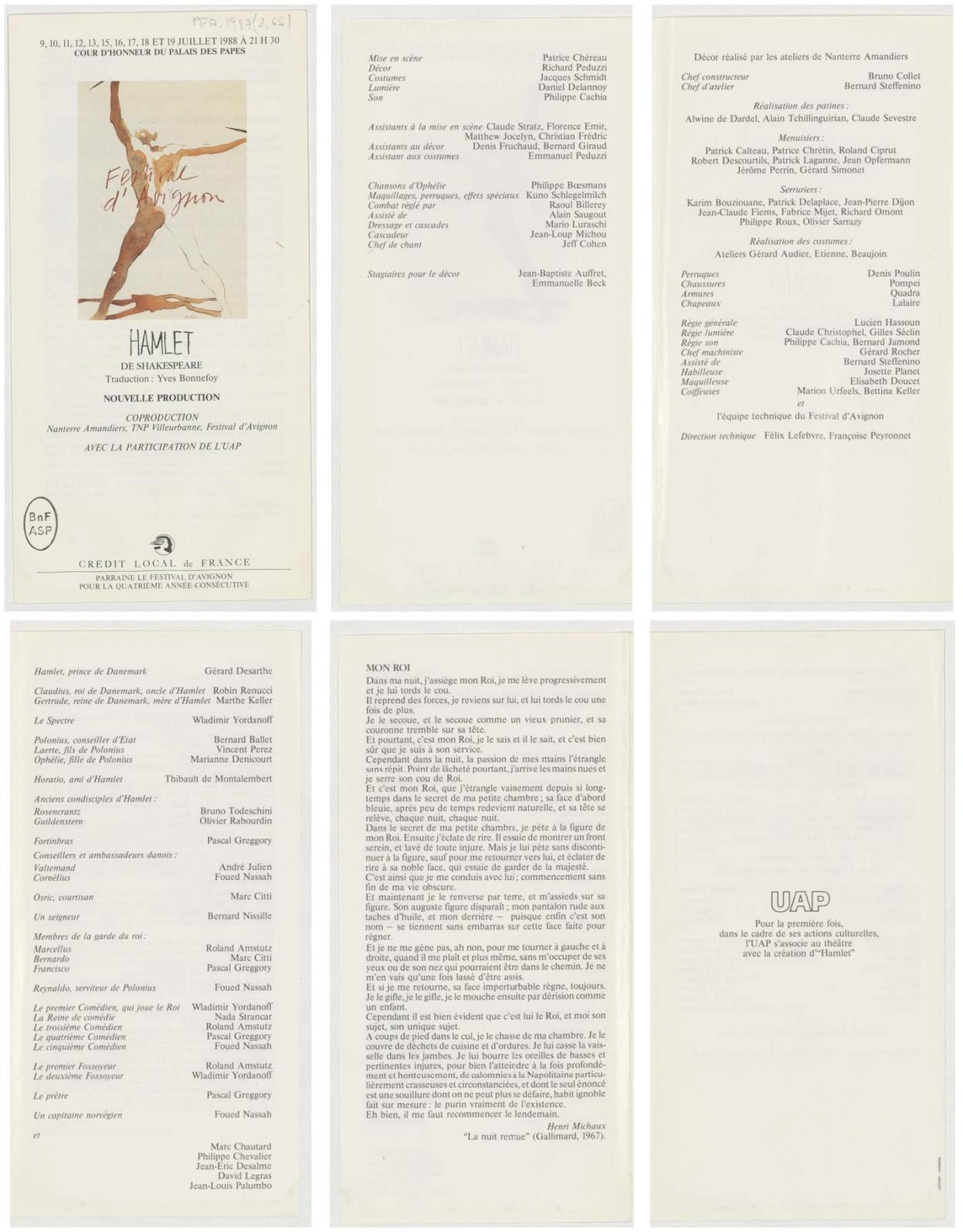}
	\caption{The show program (6 pages) of Patrice Chéreau, \textit{Hamlet}, Festival d'Avignon, 1988. Bibliothèque nationale de France. Recueil. Festival d'Avignon, 1988. Documents de programmation. PFA-1988 (2,65).}
	\label{fig:hamlet-program}
\end{figure}

\begin{figure}[htbp]
	\centering
	\includegraphics[width=.9\linewidth]{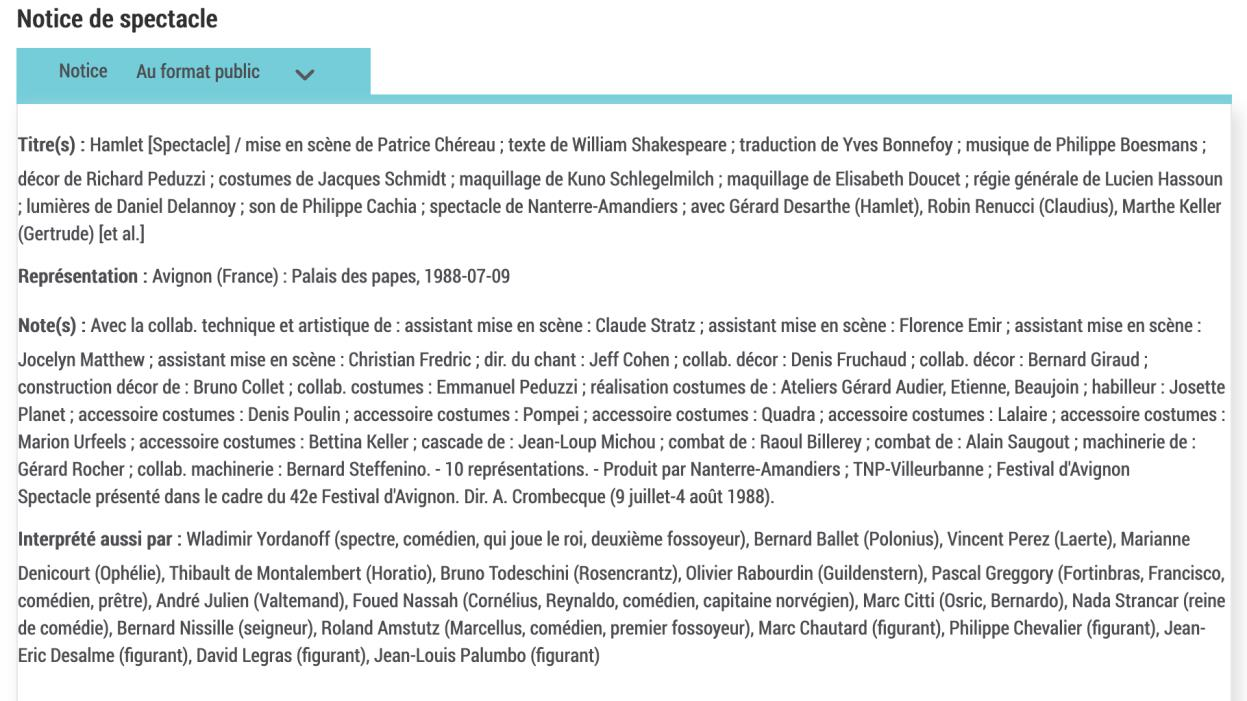}
	\caption{Screenshot of the entry for Patrice Chéreau, Hamlet, Festival d'Avignon, 1988, on the Bibliothèque nationale de France website (accessed April 29, 2026 at \protect\url{https://catalogue.bnf.fr/ark:/12148/cb39497702g}).}
	\label{fig:bnf}
\end{figure}

\begin{figure}[htbp]
	\centering
	\includegraphics[width=.9\linewidth]{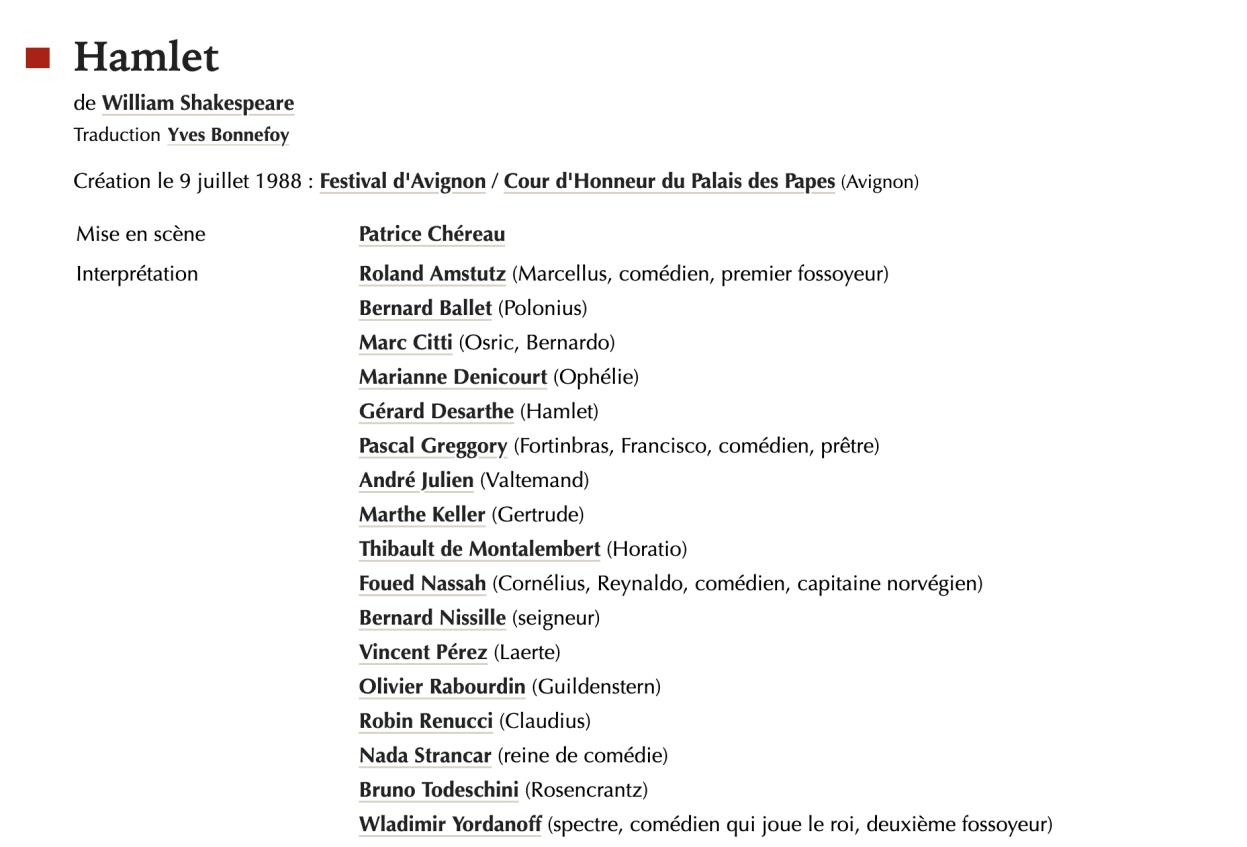}
	\caption{Screenshot of the entry for Patrice Chéreau, Hamlet, Festival d'Avignon, 1988, on the Les Archives du Spectacle website (accessed April 29, 2026 at \protect\url{https://lesarchivesduspectacle.net/s/4270-Hamlet}).}
	\label{fig:archives}
\end{figure}

\begin{figure}[htbp]
	\centering
	\includegraphics[width=.9\linewidth]{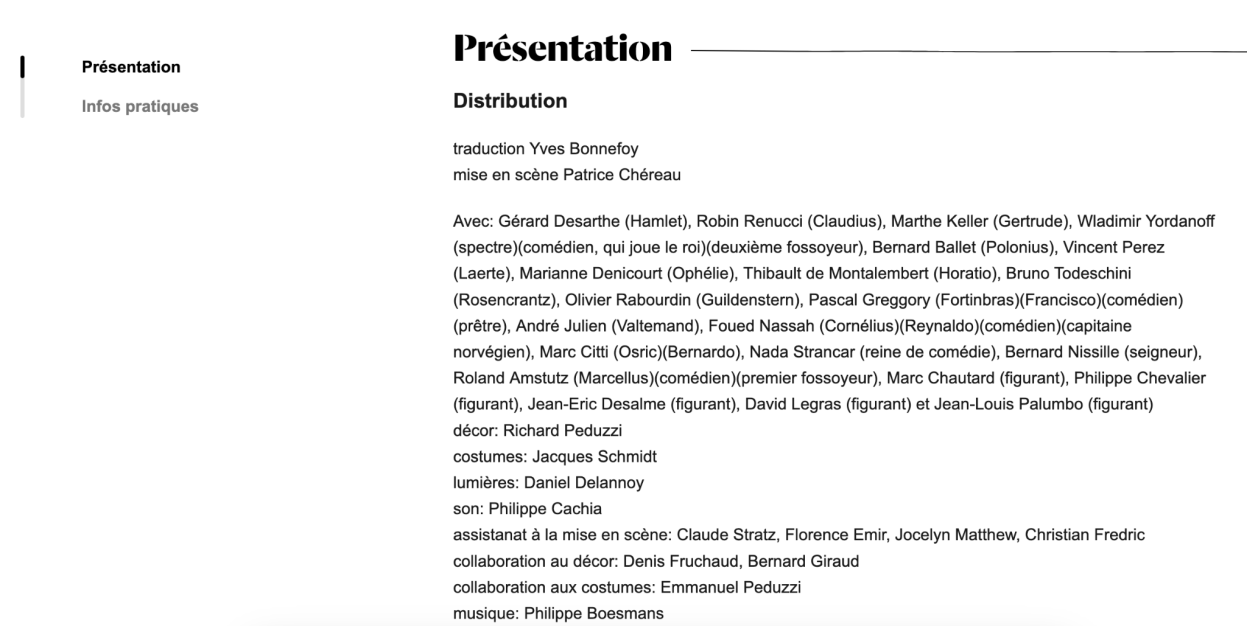}
	\caption{Screenshot of the entry for Patrice Chéreau, Hamlet, Festival d'Avignon, 1988, on the Festival d'Avignon website (accessed April 29, 2026 at \protect\url{https://festival-avignon.com/fr/edition-1988/programmation/hamlet-32322}).}
	\label{fig:festival}
\end{figure}

The 1988 production of \textit{Hamlet} directed by Patrice Chéreau at the Cour d'Honneur du Palais des Papes (Figure~\ref{fig:hamlet-program}) serves as a running example throughout this paper, illustrating both the richness of the source material and the limitations of existing databases to represent it fully. Like the other performances at the Festival d'Avignon, it is already documented in three databases: the BnF catalog,\footnote{\url{https://catalogue.bnf.fr/ark:/12148/cb39497702g}} Les Archives du Spectacle,\footnote{\url{https://lesarchivesduspectacle.net/s/4270-Hamlet}} and the Festival d'Avignon's website\footnote{\url{https://festival-avignon.com/fr/edition-1988/programmation/hamlet-32322}} (Figures~\ref{fig:bnf}, \ref{fig:archives}, \ref{fig:festival}). A comparison with the original six-page program demonstrates the systematic and compounding informational losses we aim to address. The festival's website records only the main roles; the BnF introduces semantic drift in role designations as well as omissions; Les Archives du Spectacle reproduces these omissions while mapping roles to its own undocumented controlled vocabulary, creating a chain of terminological transformations in which the original role credits disappear. Across all three sources, technical personnel are absent despite being explicitly credited in the program, as is Henri Michaux's poem that Chéreau selected as his sole note of artistic intent. The program itself remains the only complete and reliable source; recovering its information at scale is the challenge this paper addresses.

\section{State of the art}

\subsection{Automated Transcription of Heritage Documents: From OCR to VLMs}

The Festival d'Avignon programs present a particularly demanding case for automated transcription. Spanning eight decades (1947--present), the whole corpus reflects numerous typographic and layout changes ranging from minimalist single-sheet formats to elaborate multicolumn and folded booklets. Some editions incorporate decorative typefaces that blur the boundary between print and handwriting, a known challenge for both traditional OCR and VLM-based approaches.\footfullcite{kiesslingDoesContextMatter2024}
Beyond typography, the progressive internationalization of the festival adds another layer of complexity, with the addition of multilingual elements.

Over the last decade, document processing has shifted from character-level OCR pipelines, which identify individual characters through pattern matching, to end-to-end VLMs, neural systems capable of processing a document image as a whole and extracting not only text but also semantic structure and spatial relationships.\footfullcite{huangLayoutLMv3PretrainingDocument2022, liTrOCRTransformerbasedOptical2022}
Unlike traditional OCR, which requires separate preprocessing steps and struggles with irregular layouts, VLMs can handle complex, heterogeneous documents without document-specific training. They can detect hierarchical structure in the text (e.g., differentiating titles from body text), recompose paragraphs that span multiple columns, recompose words broken by line-jumps, and so on. Recent evaluations on Latin-alphabet historical prints\footfullcite{greifMultimodalLLMsOCR2025}
and on handwritten historical records\footfullcite{kimEarlyEvidenceHow2025}
demonstrate a clear advantage of VLMs (such as Gemini 2.0 Flash, Claude Sonnet 3.5, and GPT-4o) over traditional approaches (such as Transkribus Text Titan I, Print M1, Tesseract, EasyOCR, Keras, Pytesseract, and TrOCR), confirming that VLMs outperform conventional OCR and Handwritten Text Recognition methods across standard metrics.

\subsection{Ontological Frameworks for Performing Arts}

Several domain-specific initiatives have modeled performing arts data, from national aggregators such as AusStage\footfullcite{bollenDataModelsTheatre2016}
and ArtsData\footfullcite{ArtsdataDocumentation}
to ontology-based frameworks such as CapData opéra\footfullcite{peyreCapDataOperaFaciliter2024}
or the Swiss Performing Arts Data Model.\footfullcite{SwissPerformingArts}
While each addresses particular archival or disciplinary needs, none provides a generic, cross-institutional model grounded in an established standard. Therefore, to create a data model semantically expressive enough to support data capture and retrieval across diverse countries, institutions, and collections, we propose a new performing arts ontology extending a CIDOC conceptual reference model (CRM) and Linked Art.

Initially built for the museum sector and later expanded to encompass intangible cultural heritage, CIDOC CRM is the dominant reference framework for this domain.\footfullcite{lietoConceptualModelsIntangible2014}
Linked Art is a community-driven application profile of CIDOC CRM with a lower level of complexity, focused on a specific domain. The problem is that it currently lacks patterns designed specifically for the performing arts. For instance, it does not model staging roles, audiences, or the complex event hierarchies (e.g., season, production, or performance) that characterize shows. Our work proposes a targeted extension, the Linked Art Performing Arts ontology (LA-PA), to fill this gap, adding new classes, properties, and controlled vocabularies while preserving compatibility with the Linked Art core.

\subsection{Structured Information Extraction from Historical Documents}

The broader challenge of extracting structured information from unstructured or semi-structured historical documents has attracted growing attention in natural language processing (NLP) and digital humanities. Named entity recognition (NER), the automated identification and classification of persons, places, organizations, and dates in text, has emerged as a foundational technique for this task.\footfullcite{ehrmannNamedEntityRecognition2023, tualBenchmarkNestedNamed2023}

In the context of growing attention to the use of LLMs for structured knowledge extraction and the conversion of unstructured text into resource description framework (RDF) triples conforming to a domain ontology, Haonan Bian argues that two broad paradigms have emerged.\footfullcite{bianLLMempoweredKnowledgeGraph2025}
The first, \textit{schema-free}, creates structure dynamically from text, building its own network of concepts without any predefined framework (e.g., GraphRAG or retrieval-augmented generation).\footfullcite{edgeLocalGlobalGraph2025}
While these methods can surface unexpected connections, they offer no guarantee that the concepts produced will be consistent across documents or that they will align with the controlled vocabularies and classification standards that heritage institutions rely on. The second paradigm, \textit{schema-based}, uses a predefined ontology as a structural backbone that constrains LLM generation from the outset. Recent surveys confirm that this approach yields higher consistency and reduced hallucination rates compared to unconstrained generation.

Within the digital humanities, a small number of projects have explored hybrid extraction pipelines on historical corpora. To our knowledge, no automated pipeline has been developed specifically for the extraction of structured information from performing arts programs. The closest existing work to our own is the project by Matteo Romanello et al.,\footfullcite{romanelloPredictingFictionalTime2025}
who deploy a hybrid pipeline combining NER, entity linking, and LLMs to extract spatio-temporal features from approximately 600 17th-century French theater plays sourced from Gallica. Their evaluation, benchmarking nine models from Phi-4-mini to Claude Sonnet 3.7, confirms both the viability of LLM-based structured extraction on theatrical corpora and its limits: Claude Sonnet 3.7 achieves the highest accuracy overall, but all tested models systematically hallucinate Wikidata QIDs for fictional locations, even when the location string itself is correctly predicted. This finding directly motivates our approach of constraining generation through a domain ontology instead of relying on open-ended LLM output.

\section{Methodology}

\begin{figure}[htbp]
	\centering
	\includegraphics[width=\linewidth]{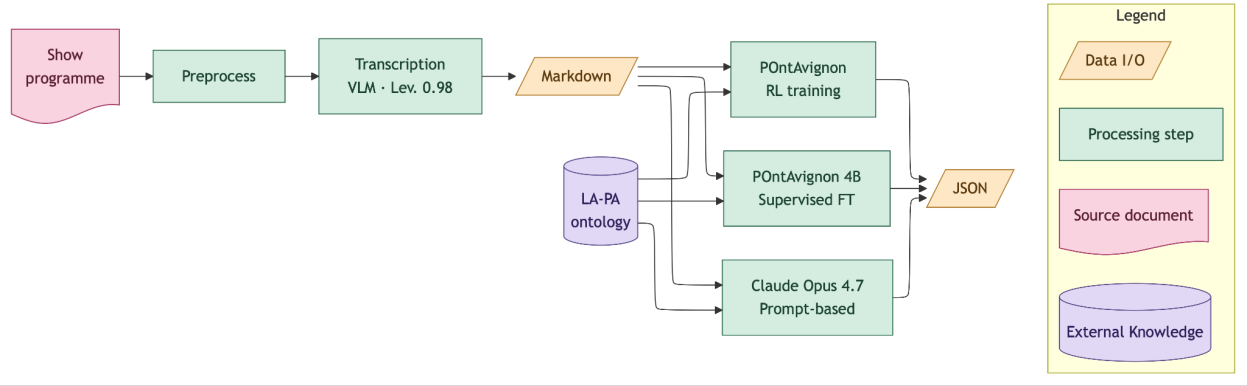}
	\caption{Overview of the three-stage pipeline: VLM transcription, ontological modeling (LA-PA), and semantic annotation.}
	\label{fig:pipeline}
\end{figure}

The pipeline\footnote{The pipeline, the ground truth, the evaluation code, and all the results discussed in this section are available at \url{https://github.com/stage-to-data/programmes-to-data-example}.} operates in three sequential stages (Figure~\ref{fig:pipeline}): (1) document transcription via VLMs, (2) ontological modeling through LA-PA, and (3) semantic annotation via different approaches. Each is described in turn in the following sections. Transforming heritage documents into structured data requires evaluating two distinct aspects of quality: transcription accuracy and semantic annotation accuracy. For both, we construct a manually verified ground truth drawn from the Festival d'Avignon programs, covering the full typographic and layout diversity of the corpus.

\subsection{Document Transcription via VLMs}

The first step consists of transcribing the decomposed programs using a VLM. This section details each process step, our ground truth creation for evaluation, our prompt engineering approach, and the advantages and caveats of using VLMs for transcription.

To evaluate different prompts and models, we created a ground truth corpus comprising 30 show programs representing digitized and born-digital documents, including challenging scenarios (e.g., upside-down text, poor quality scans). The corpus contains 151 pages, 37,466 words, and 256,585 characters. With initial VLM assistance, we manually annotate each page in markdown format, differentiating only title (irrespective of hierarchical level) and non-title text. The transcriptions are available on GitHub.\footnote{The figures given at \url{https://github.com/stage-to-data/programmes-to-data-example/tree/main/evaluation/ground-truth} correspond to commit \texttt{c7fc5733} of the upstream corpus. The corpus is revised over time and figures scored against different states are not comparable, so every run in the repository records the commit it was scored against.}

Using our ground truth for evaluation, we tested different models (OpenAI's GPT-4o, Anthropic's Claude Sonnet 3.7, and various llama-vision versions run locally via Ollama) to engineer an optimal prompt. Because precise NER is more crucial than strict section ordering in our application, standard document-wide word error rate (WER)/character error rate (CER) calculations proved inadequate. Instead, our evaluation notebook\footnote{See \url{https://github.com/stage-to-data/programmes-to-data-example/blob/main/evaluation/evaluation.ipynb}.} weighted word counts and calculated metrics across multiple granularities: file-level Levenshtein distance; sentence/line-level precision, and recall, obtained by first matching units between the two texts and then computing WER, CER, and Levenshtein on the matched pairs. The same procedure was applied to named entities. Results are compiled into browsable markdown files.\footnote{At \url{https://github.com/stage-to-data/programmes-to-data-example/tree/main/evaluation/results}, there is one folder per run, grouped by campaign. Metrics for every run in a single table are available at \url{https://github.com/stage-to-data/programmes-to-data-example/blob/main/evaluation/results/corrected-metrics.md}.}

\begin{table}[htbp]
	\caption{Results of two of the different test configurations, scored against the same state of the ground truth. Ground truth commit \texttt{c7fc5733}. All figures are weighted by page word count. P/R decomp.\ is precision and recall at the sentence/line level; Lev.\ decomp.\ is the mean Levenshtein ratio over matched units.}
	\label{tab:results}
	\centering
	\small
	\begin{tblr}{
			width   = \textwidth,
			colspec = {Q[l] X[c] X[c] X[c] X[c] X[c]},
			row{1}  = {font = \bfseries, valign = m},
			cells   = {valign = m},
		}
		\toprule
		Configuration
		& {Lev. full\\document}
		& {P/R\\decomp.}
		& {Lev.\\decomp.}
		& {P/R NER\\decomp.}
		& {Lev. NER\\decomp.} \\
		\midrule
		Prompt 1 Claude & 0.9817 & 0.79/0.84 & 0.9856 & 0.94/0.91 & 0.9954 \\
		Prompt 4 Claude & 0.9818 & 0.82/0.86 & 0.9868 & 0.94/0.91 & 0.9958 \\
		\bottomrule
	\end{tblr}
\end{table}

Table~\ref{tab:results} summarizes two of the prompt configurations we tested\footnote{The full set of prompts tested is available at \url{https://github.com/stage-to-data/programmes-to-data-example/tree/main/evaluation/prompts}, and the complete results for each pass, including the two intermediate prompts, are available at \url{https://github.com/stage-to-data/programmes-to-data-example/tree/main/evaluation/results/final-tests}. The gain from prompt 1 to prompt 4 lies in coverage rather than in fidelity: precision at the decomposed level rises by more than three points, while the Levenshtein ratios, already above 0.98, barely move. This asymmetry is informative in itself: once units are matched, what the model transcribes is transcribed almost exactly. What distinguishes prompts is how much of the page they capture at all. An earlier round of testing, scored against a previous state of the ground truth, gave a document-wide Levenshtein ratio of 0.9833 and NER precision/recall of 0.94/0.91 for prompt 1 with Claude Sonnet 3.7, against 0.9699 and 0.85/0.87 for GPT-4o. We did not re-run GPT-4o on the revised ground truth, so the two are reported separately rather than in the same table. The detailed results of our final prompt are at \url{https://github.com/stage-to-data/programmes-to-data-example/blob/main/evaluation/results/final-tests/prompt-4/scores.md}. The markdown files for the Chéreau example are available at \url{https://github.com/stage-to-data/programmes-to-data-example/blob/main/evaluation/results/final-tests/prompt-4/scores.md}.} with encouraging results: document-wide Levenshtein ratio (word count--weighted) of 0.98, and a mean named entity Levenshtein ratio\footnote{Named entity Levenshtein ratios are calculated for the content of each detected named entity pair between test and ground truth.} of 0.99 on matched entities. Detailed information on our final chosen prompt and model is available in Appendix 1. After finalizing the model and prompt, we processed our entire corpus.

\begin{figure}[!htb]
	\caption{First page of the \textit{Hamlet} program by Patrice Chéreau in 1988, showing its original scan, VLM transcription, and EasyOCR transcription. Transcription errors are highlighted in yellow; no errors were identified in the VLM output.}
	\label{fig:transcription}
	\centering
	\footnotesize
	\begin{tblr}{
			width   = \textwidth,
			colspec = {X X X},
			cells   = {halign = l, valign = t},
			hlines, vlines,
		}
		\includegraphics[width=\linewidth,valign=T]{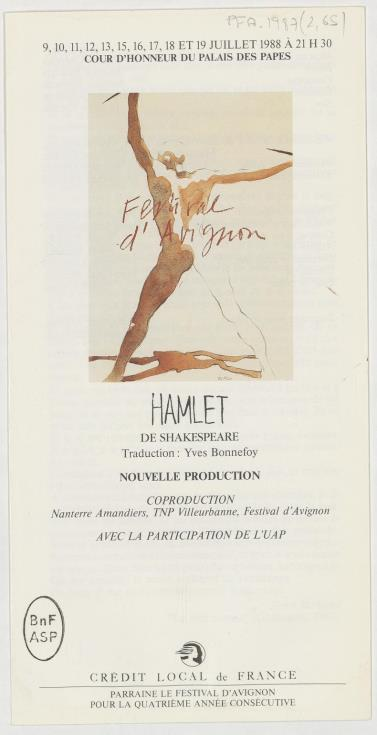}
		&
		PFA.1987(2,65)
		\par\bigskip
		9, 10, 11, 12, 13, 15, 16, 17, 18 ET 19 JUILLET 1988 À 21 H 30
		\par\smallskip
		COUR D'HONNEUR DU PALAIS DES PAPES
		\par\bigskip
		Festival
		\par\smallskip
		d'Avignon
		\par\bigskip
		\# HAMLET
		\par\bigskip
		DE SHAKESPEARE
		\par\smallskip
		Traduction : Yves Bonnefoy
		\par\bigskip
		NOUVELLE PRODUCTION
		\par\bigskip
		COPRODUCTION
		\par\smallskip
		Nanterre Amandiers, TNP Villeurbanne, Festival d'Avignon
		\par\bigskip
		AVEC LA PARTICIPATION DE L'UAP
		\par\bigskip
		BnF
		\par\smallskip
		ASP
		\par\bigskip
		CRÉDIT LOCAL de FRANCE
		\par\bigskip
		PARRAINE LE FESTIVAL D'AVIGNON
		\par\smallskip
		POUR LA QUATRIÈME ANNÉE CONSÉCUTIVE
		&
		\surligne{FFA,17 89(2, \textless s} 9,10, 11, 12, 13, 15, 16,17, 18 ET 19
		JUILLET 1988 À 21 H 30 COUR D'HONNEUR DU PALAIS DES PAPES
		\par\smallskip
		\surligne{F d'}
		\par\smallskip
		\surligne{Yhor}
		\par\smallskip
		HAMLET DE SHAKESPEARE Traductio\surligne{n }Yves Bonnefoy NOUVELLE PRODUCTION
		\par\smallskip
		COPRODUCTION
		\par\smallskip
		Nanterre Amandiers\surligne{;} TNP Villeurbanne\surligne{;} Festival
		\surligne{dA}vignon AVEC LA PARTICIPATION DE L'UAP
		\par\smallskip
		BnF AS\surligne{F}
		\par\smallskip
		CRED\surligne{I T} LOCAL de FRANCE PARRAINE LE FESTIVAL D'AVIGNON POUR LA
		QUATRIÈME ANNÉE CONSÉCUTIVE
		\\
	\end{tblr}
\end{figure}

Claude's performance is impressive: results confirm the superiority of VLMs over standard OCR, as shown in the Hamlet program by Patrice Chéreau example (figure~\ref{fig:transcription}). EasyOCR achieves a median WER of 0.20 on born-digital documents, whereas Claude Sonnet 3.7 achieves a document-wide weighted Levenshtein ratio of 0.98 and a mean named entity Levenshtein ratio of 0.99. Processing the full corpus cost approximately \$80, with total project costs (including testing and evaluation) around \$100.

\subsection{Ontological Modeling: The Linked Art Performing Arts Extension}

Multimodal transcription produces source-faithful markdown text from program pages. The next challenge is representational. How do we move from the richly digitized source material to a faithful representation of the historical facts captured in this data that is generally understandable and reusable by the performing arts community? Part of this challenge is ontology engineering: how do we represent roles, names, venues, and dates in machine-readable representations aligned with a shared ontological framework?\footfullcite{brusekerSemanticReferenceData2025}
The other part is communication: how do we represent and communicate this model in a way that overcomes the challenges of a rich ontology such as CIDOC CRM or an application profile such as Linked Art?

Rather than propose yet another institution-specific schema, we developed LA-PA, a domain extension built on Linked Art and released in April 2026 as an openly documented standard (v0.9). LA-PA is developed collaboratively by an international working group made up of researchers, ontologists, and cultural heritage professionals representing multiple institutions and areas of performing arts expertise. LA-PA builds itself up in an ecosystem of extant maintained ontologies from which appropriate classes and properties can be selected and deployed, including CIDOC CRM, Art and Architectural Argumentation Ontology (AAAo), Power, War, Religion and Relations Ontology (PWRo), Creative Processes Representation Ontology (CPRo), and CRM Digital (CRMdig). Regarding controlled terminology, we link to existing standards where possible. These include Library of Congress Code List for Relators and BnF Roles for professions; ISO 639 for languages; Getty Art and Architecture Thesaurus (AAT) for event types; and VIAF, ISNI, and Wikidata for external identifier alignment.

\begin{figure}[htbp]
	\centering
	\includegraphics[width=\linewidth]{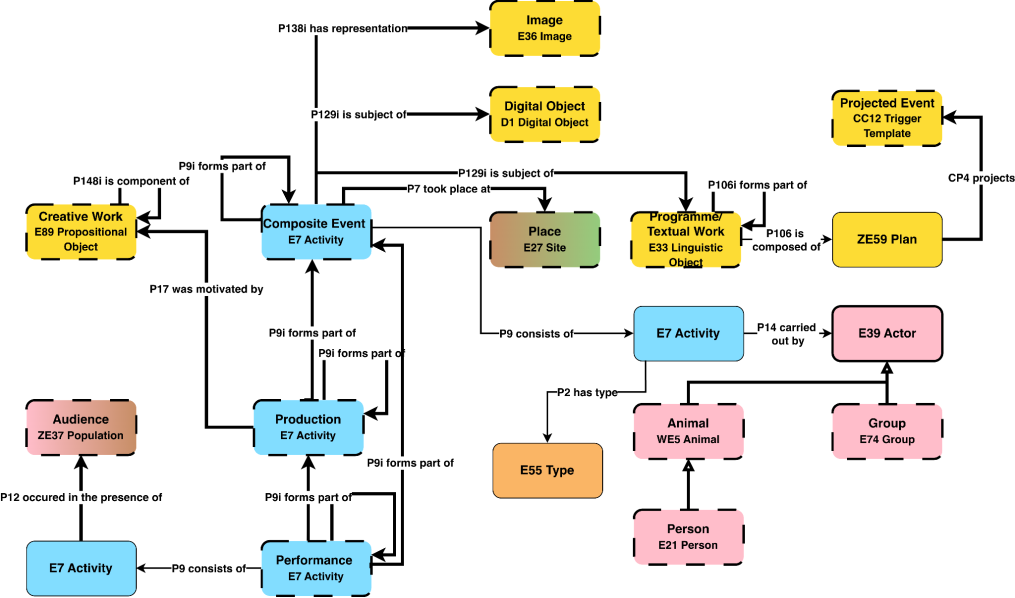}
	\caption{Overview of the Linked Art Performing Arts ontology (LA-PA).}
	\label{fig:lapa}
\end{figure}

LA-PA consists of 13 main semantic models, organized so as to facilitate the accurate capture and representation of the main entities of interest in performing arts documentation. We can group these models epistemically into three groups: events, objects, and agents (figure~\ref{fig:lapa}).

The event models provide a suite of semantic data models that guide the creation and interpretation of performing arts data and are capable of capturing the main kinds of events of interest to distinguish, document, and interrelate in this field of study. These are the following:

\begin{itemize}
	\item The Composite Event (\texttt{crm:E7\_Activity | aat:300460397}) models a bounded container event, a festival edition or theater season, within which multiple productions are programmed. In our running example, this corresponds to the 42nd edition of the Festival d'Avignon (1988).
	\item The Production (\texttt{crm:E7\_Activity | aat:300460399}) documents a realization of a Work (see the object section) in a specific context, recording creative team, producing company, funders, venue, and overall timespan; for instance, Chéreau's staging of Hamlet at the Cour d'honneur du Palais des Papes, Festival d'Avignon, July 9--19, 1988.
	\item The Performance (\texttt{crm:E7\_Activity | aat:300069200}) represents a single show on a given date, capturing time-specific details and last-minute cast variations; for instance, the world premiere of Hamlet, July 9, 1988. Composite Event, Production and Performance are instances of the Linked Art Activity class, differentiated through Getty AAT vocabulary terms.
	\item The Projected Event (\texttt{cpro:C12\_Event\_Trigger\_Template}) is a distinctive feature of LA-PA: it models the event as announced, typically as it appears in a printed program, as opposed to what actually occurred. The distinction is important for source-critical historical research: what was announced might differ from what was performed (due to cast replacements, cancellations, or additions). This class is indispensable for working with program data, where the document records intention rather than fact.
\end{itemize}

These models work together with the object and agent layers. The object models consist of Creative Work, Programme / Textual Work, and Place, which notably differ the most from Linked Art, as well as Image and Digital Object.

\begin{itemize}
	\item The Creative Work (\texttt{crm:E89\_PropositionalObject}) is fundamental to the modeling, representing the abstract artistic conception prior to material realization: not the play text, but the director's staging project. A single Creative Work may give rise to multiple Productions across different venues and contexts while remaining recognizably the same artistic entity. Chéreau's staging of Hamlet was realized as different Productions (e.g., a premiere in Avignon and a tour in different cities like Milan and Paris) without ceasing to be the same Creative Work.
	\item The Programme / Textual Work model (\texttt{crm:E33\_Linguistic\_Object}) is the model for the documentation of the chief source of data itself: the program. While the program is in many essential aspects a document like other documents aligning with the Textual Work model of Linked Art, its characteristic also as a plan for a potential event which may not be realized motivates special modeling linking it to the Projected Event model, for which it provides the historical data of intended, but not necessarily realized, events.
	\item The Place model (\texttt{crm:E27\_Site}) supports the documentation of the places at which performing arts events occurred. This model diverges from the modeling of place in Linked Art as a mere abstract geometric entity. Instead, motivated by the materiality of venues and the importance of the study of the materiality and historicity of places, it is modeled as a physical site with its own potential topographic as well as historiographic connections.
\end{itemize}

Lastly, the agent layer deploys four models: person and group, which follow standard Linked Art, plus animal and audience.

\begin{itemize}
	\item The Audience model (\texttt{aaao:ZE37\_Population}) enables the documentation of ad hoc groups of individuals present at performances. Such groups are not characterized by collective agency but instead are bound together by their presence at a moment. Nevertheless, the study of their composition, size, and so on is made possible through this model, going beyond the traditional documentation of memory institutions.
	\item The Animal model (\texttt{pwro:WE5\_Animal}) is also a unique extension to the Linked Art modeling paradigm driven empirically from the documentation of the activity of animals in performance events.
\end{itemize}

\begin{figure}[htbp]
	\centering
	\includegraphics[width=\linewidth]{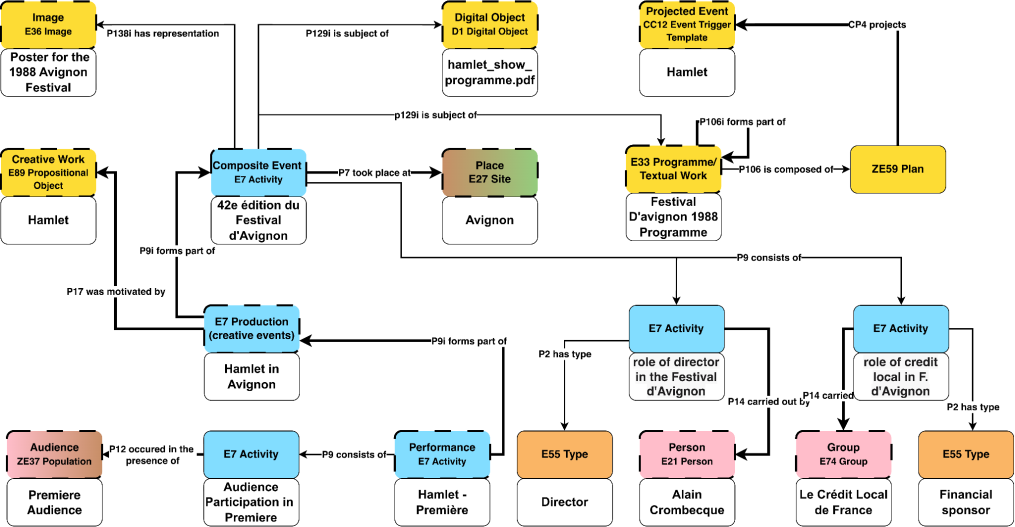}
	\caption{The model applied to the premiere of \textit{Hamlet} staged by Patrice Chéreau for the Festival d'Avignon in 1988.}
	\label{fig:lapa-hamlet}
\end{figure}

The other half of the modeling work has been to build data objects that make it tractable to its potential audiences of researchers, developers, and semantic data specialists. Real-world use cases both test and illustrate the model (Figure~\ref{fig:lapa-hamlet}). Current issues, presentations, and examples are available on GitHub.\footnote{See \url{https://github.com/stage-to-data/linked-art-pa}.} The models are given a permanent documentation format using the Pletka Semantic Pattern Library.\footnote{See \url{https://dev.pletka.io/projects/STA}.}

\subsection{Semantic Data Annotation}

Transforming transcribed program text into RDF triples aligned with LA-PA requires solving what Pascal Hitzler et al.\footfullcite{hitzlerNeuralsymbolicIntegrationSemantic2020}
call the ``knowledge acquisition bottleneck'': manual encoding of 1,840 programs encompassing thousands of staging events and associated entities would take months. Over the last year, research progress in LLMs has been largely driven by the development of ``reasoning'' or ``thinking'' models. All state-of-the-art models currently include a preliminary step (the ``draft'') that allows the models to process more complex problems and/or perform intermediary actions. The emerging approach of training dedicated reasoning models via reinforcement learning, a technique whereby a model learns from the correctness of its outputs rather than from labeled examples, pioneered for verifiable domains such as mathematics and programming,\footfullcite{guoDeepSeekR1IncentivizesReasoning2025}
opens new possibilities for the semi-verifiable task of ontology-constrained semantic annotation. We pursued two successive automated approaches before converging on the current production pipeline.

\begin{figure}[htbp]
	\centering
	\includegraphics[width=0.95\linewidth]{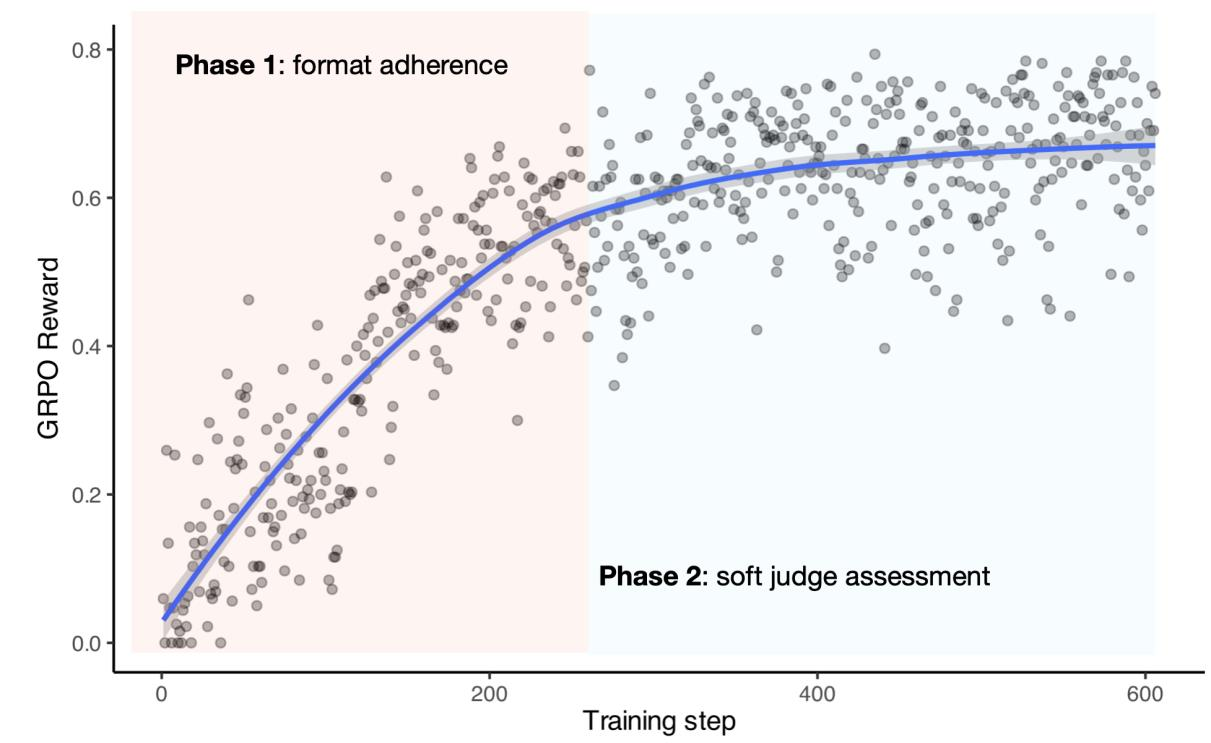}
	\caption{Training curve of POntAvignon (0.4B), showing two successive phases: formal constraint learning (phase 1) and reasoning-intensive semantic refinement (phase 2).}
	\label{fig:training}
\end{figure}

The initial strategy, built on a first draft of the ontology, adapted Pleias-350m, a small language model mid-trained on Wikidata with generalist text-to-knowledge-graph capabilities.\footnote{Since February 2025, Pleias has partnered with the Wikimedia Foundation to develop ethical models on data from the Wikimedia project, which led to the training of a ``Wikidata'' reasoning model: a 350 million transformer able to map any unstructured text in the main European languages into a Wikidata structured item, using systematically the 4,000 main properties developed over a decade by the community. The model was trained using a methodology of ``mid-training'' at Jean Zay, the leading GPU public cluster in France, combining 10 million samples of Wikidata items with synthetic text as input.} Domain adaptation used online reinforcement learning with Group Relative Policy Optimization (GRPO): at each step, the model received program excerpts as input and generated eight alternative structured drafts. A formal reward enforced strict compliance with the predefined LA-PA property set: any draft containing out-of-vocabulary properties received a zero reward. Drafts passing this filter were then evaluated by a judge model (Gemma 12b) for semantic accuracy against ground truth, on a 0 to 10 scale rewarding partial success (soft reward). The training curve (Figure~\ref{fig:training}), combining formal and soft rewards, confirmed the feasibility of the approach. A first phase of aggressive constraint learning, where the model rapidly abandoned out-of-vocabulary properties, was followed by a more gradual second phase of reasoning-intensive semantic refinement. Reasoning traces contribute to increased accuracy of the model and provide some level of explainability. Due to being built on top of a model primarily trained on Wikidata, the ontology is immediately compliant with Wikidata, and, as shown in Appendix 2, the reasoning traces explicitly call for the Wikidata property IDs. The resulting model, POntAvignon (0.4B parameters), was released on Hugging Face.\footnote{See \url{https://huggingface.co/LLMDH/POntAvignon}.}

Two structural limitations emerged in practice. First, the model was deliberately trained on single-show submissions: multi-show programs produced either partial extraction or entity conflation. Second, the pretraining context window of Pleias-350m was too small to consider the complete program, a critical constraint given that Avignon programs regularly span six to ten pages with detailed cast and crew listings. The output consequently covered only a subset of LA-PA entities, and cross-referencing between models (e.g., linking a Production to its Composite Event or its source Programme / Textual Work) was not supported.

The second iteration, POntAvignon-4b,\footnote{See \url{https://huggingface.co/Pclanglais/POntAvignon-4b}.} addressed the scale constraint by building on Qwen3-4B (4 billion parameters), which offers a 16k-token context window. Rather than reinforcement learning, this version uses full supervised fine-tuning (SFT; 3 epochs, bf16). Reasoning traces, structured in the SYNTH format developed by Pleias, were generated for a seed set of 3,110 samples by Claude Sonnet and then scaled to the full dataset using a Gemma 12b backreasoning model trained on these traces. The model decomposes extraction into seven sequential entity queries corresponding to the main LA-PA layers: the Creative Work (\texttt{A}), Production metadata (\texttt{B\_meta}), cast (\texttt{B\_cast}), crew (\texttt{B\_crew}), individual Performance (\texttt{C}), Composite Event, and source Programme / Textual Work. The result of the training process was that on held-out test data, it achieves 96.6\% token accuracy and 97\% valid JSON rate.

Two limitations remain. First, POntAvignon-4b was trained on a previous iteration of the ontology; the current LA-PA v0.9 introduces additional models (Projected Event, Audience, and composite program hierarchy) and substantially revised field definitions and controlled vocabulary rules that are not yet reflected in the model. This highlights the ongoing challenge of ontology development outpacing model training. Second, large show productions, those with 30 or more team members cited in the program, risk truncated outputs near the context limit, precisely the cases where completeness is most critical for historiographical purposes.

Pending the retraining of POntAvignon on LA-PA v0.9, the current production pipeline uses Claude Opus 4.8 with a structured extraction prompt encoding the full ontology.\footnote{See \url{https://github.com/stage-to-data/programmes-to-data-example/tree/main}.} The prompt defines nine LA-PA models (simplifying the 13 models of the ontology in the context of show programs from the same corpus) with their fields, instructions, controlled vocabularies (Getty AAT), ID construction rules, verbatim copy requirements, and edge-case handling (e.g., composite programs, reprise rule, multi-role agents, document vs.\ production credit separation). The nine models define the semantics of each field; the model returns a single consolidated table, one row per Production, whose 58 columns flatten those models---a format convertible to RDF or JSON-LD.

\begin{figure}[htbp]
	\centering
	\includegraphics[width=.95\linewidth]{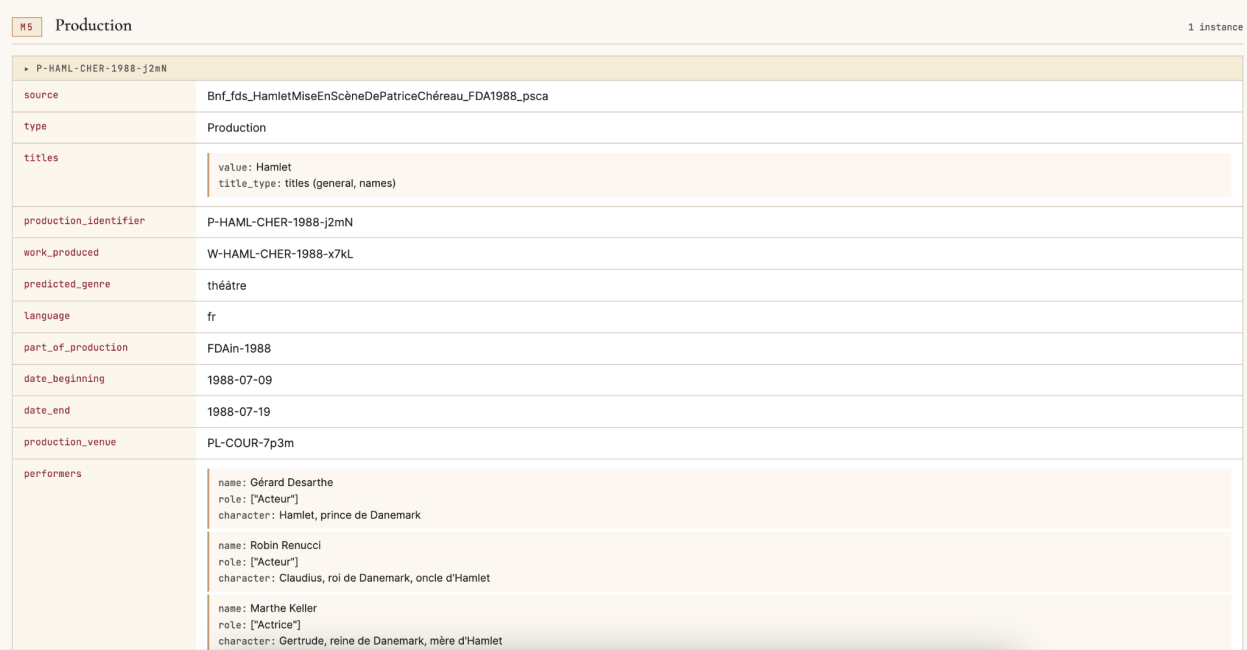}
	\caption{Screenshot of the Production model output for Patrice Chéreau's \textit{Hamlet} in 1988 rendered as a table, as generated by the Claude Opus 4.8 annotation prompt.}
	\label{fig:production}
\end{figure}

As an initial quality check, we applied the pipeline to different programs and compared the output against the manually verified reference. This helped refine our prompting. For example, for the 1988 Chéreau Hamlet program, the output covers seven models: 1 Composite Event, 1 Programme document, 1 Programme / Textual Work (Shakespeare's Hamlet, Bonnefoy translation), 1 Creative Work, 1 Production with 33 performers and 67 team members, 10 individual Performances (9--19 July 1988, 21:30), and 1 Place (figure~\ref{fig:production}).\footnote{The extracted data, the markdown submitted to the model and a note on the three approaches are available at \url{https://github.com/stage-to-data/programmes-to-data-example/tree/main/Hamlet_example/outputs/text/Bnf_fds_HamletMiseEnScèneDePatriceChéreau_FDA1988_psca/data}.}

\section{Discussion and perspectives}

Processing the Festival d'Avignon corpus with a 0.98 Levenshtein ratio for text extraction confirms the viability of the approach on heterogeneous historical documents. The primary reproducibility risk is the instability of proprietary LLM APIs: model updates may alter transcription behavior, and outputs are nondeterministic by design. Mitigation strategies include version pinning, systematic evaluation against fixed ground truth, and the parallel development of open-source VLM alternatives.

To achieve a balance between standardization, fidelity, and understandability, Linked Art's data patterns were used and extended with terminology aimed at the performing arts community. Two persistent modeling challenges emerged in this process. First, the actor--character relationship lacked sufficient modeling: neither CIDOC CRM nor Linked Art nor LRMoo provides a semantic pattern for a performer's portrayal of a fictional character, requiring an original solution in the LA-PA extension. Second, there was no sufficient standardized list of job titles: the performing arts deploy a far richer vocabulary of functions than the visual arts domains for which Linked Art was originally designed. Getty AAT coverage proved insufficient, requiring complementary alignment with BnF and Library of Congress role vocabularies and the development of project-specific controlled vocabularies for roles and event types.

Across the three approaches, a consistent finding emerged: constraining LLM generation through an explicit ontological framework reduced hallucination rates, regardless of whether the constraint was implemented as a reward policy, a fine-tuning objective, or a prompt-encoded schema. This positions semantic annotation as a productive hybrid between fully verifiable tasks, like mathematics or code where automated evaluation is relatively straightforward, and purely subjective ones, where it is impossible. Ontological compliance is semi-verifiable: whereas properties are either in the defined set or not, semantic accuracy is more partial. It is precisely this characteristic that makes LLMs ideally suited for the task, both in the hard signals of reinforcement learning and softer semantic rewards, or when used as few-shot learners that do not require an exhaustive ground truth.

A complementary pipeline under development within the STAGE project approaches the same extraction task through yet another architectural logic, called an ``agent execution harness,'' or software guiding the LLM's output. This approach decomposes extraction into a sequence of explicit, observable steps, each with recorded provenance, and separates model inference from ontology management. This mirrors a growing view within research into AI systems that harness design is key for ensuring model reliability and maximizing model performance on a given task, as illustrated by a recent focus on coding harnesses like Anthropic's Claude Code and OpenAI's Codex.\footfullcite{AgentHarnessLarge}
This separation matters practically: as LA-PA continues to evolve, ontology updates and vocabulary expansions can be incorporated without energy-intensive retraining.

Producing structured data is only the first stage of the workflow. The next phase concerns the deployment of the derived data in a queryable, publicly accessible infrastructure. The structured RDF output will be published via a Linked Art--compliant endpoint, with entity reconciliation against VIAF, ISNI, and Wikidata completed prior to release. A search and exploration interface designed for both researchers and heritage professionals is under development, with the aim of making the Festival d'Avignon data accessible without requiring SPARQL expertise. This implementation phase will also serve as a test case for the data governance framework developed within STAGE, addressing questions of versioning and long-term preservation.

\section{Conclusion}

Our workflow demonstrates the potential of combining LLMs with domain-specific ontologies to transform cultural heritage materials into structured data while preserving textual integrity and semantic richness. Using Anthropic's Claude model for transcription and an ontology-constrained extraction pipeline for semantic annotation, we show that constraining LLM generation through a knowledge graph effectively limits hallucinations while maintaining content accuracy. The result is the scalable creation of literal transcriptions of source documents alongside normalized, documented data aligned with external knowledge graphs. Beyond this technical contribution, our findings suggest that working with LLMs forces a productive rethinking of research questions in terms of token cost, model intelligence, and now also harness design.

The Festival d'Avignon corpus is a proof of concept for a broader ambition of generating \textit{catalogues raisonnés} for performing arts creators: comprehensive, chronologically ordered records of all productions associated with a director, choreographer, or ensemble. Queries of the form ``list all productions directed by X, with venues, cast, and dates'' become straightforward. This scholarly genre, well established in art history, has no equivalent in the performing arts, where the dispersal of documentation across institutional collections makes such assembly effectively inaccessible to individual researchers. Such catalogs would renew performing arts historiography in at least three ways. First, they would make touring visible as a research object, a structurally central performing arts phenomenon whose documentation is necessarily dispersed. Second, they would enable quantitative approaches to career analysis and institutional network mapping currently limited to small, manually assembled datasets. Third, they would provide a shared factual infrastructure on which interpretive scholarship can build, analogous to the function that \textit{catalogues raisonnés} have long served in art history.

Realizing this potential requires three parallel developments: systematic entity reconciliation with authority files; implementation of the derived data in a queryable, publicly accessible infrastructure; and institutional partnerships for cross-corpus scaling beyond Avignon. The Festival d'Avignon corpus provides both a first large-scale demonstration of the new research agenda that structured program data make possible and a foundation for the collaborative construction of the interoperable infrastructure that would transform performing arts history into a field capable of large-scale, data-driven inquiry.

\section{Funding acknlowedgment}

Funded by the European Union (ERC, STAGE, grant agreement no.\ 101097091). Views and opinions expressed are however those of the authors only and do not necessarily reflect those of the European Union or the European Research Council. Neither the European Union nor the granting authority can be held responsible for them.

\printbibliography

\clearpage
\section{Appendix 1: From PDF to Markdown (MD): Claude Prompt}

\begin{Verbatim}[breaklines=true,breaksymbolleft=,breaksymbolright=,breakanywheresymbolpre=,breakautoindent=true,fontsize=\small]
This image corresponds to a page in a theater program.
You will perform a complete and detailed OCR analysis of this image.
DO NOT generate or use any code, do not use Tesseract or pytesseract — only use your internal vision to read the image.
Extract all visible text WITHOUT changing it.
DO NOT summarize, paraphrase, or infer missing text. DO NOT invent people’s names.
Retain all spacing, punctuation, and formatting exactly as in the image.
You MUST follow the following rules when dealing with these specific cases:
- All titles are indicated with a preceding # with no hierarchy (example: # Title content). If the title contains line breaks, recompose the title into one single line.
- Exponents are given as lower case roman characters (example: XVe)
- Footnotes are given as special characters (example: word¹)
- When you detect the number of a page, return it at the start of the text using the word “PAGE” followed by the number (example: PAGE 13)
- Do not describe images or logos, simply extract any text within them.
- Do not insert special characters for things like columns and page separation
- For distribution: if you detect a table-like structure, typography variation that could be interpreted as a function/name or name/function pairing, or a string of points linking a function and name, apply the following structure:
Element 1 : Associated element 1, Associated element 2
Element 2 and Element 3, description: Associated element 3 and Associated element 4
(Note that in the case of enumeration, there are NO line breaks, there are commas)
- If you detect any other function/name or name/function pairs that are separated by a character like | or /, change the character to a : (example: element 1 : associated element 1)
- If you detect a completely empty page, or if the page is much too faint or blurry to confidently transcribe, simply return: [UNABLE TO TRANSCRIBE]
- Some pages may include texts of different languages, or old languages (such as old French). It is very important to NOT CHANGE the content, simply transcribe the letters you see.
Check the overall readability of your transcription and make sure it reflects the structure of the original document.
The output should be ONLY the transcribed content (DO NOT add comments like “this is a theatre program” or “I was unable to reliably transcribe this”).
\end{Verbatim}

\section{Appendix 2: Samples of Generation After Reinforcement Learning Training}

\begin{Verbatim}[breaklines=true,breaksymbolleft=,breaksymbolright=,breakanywheresymbolpre=,breakautoindent=true,fontsize=\small]
<|thinking_start|>
- Oh, this looks like a theater program!
- “Coquin de Coq” is clearly the title of the show
- Sean O’Casey is mentioned as “de” (meaning “by” in French) - probably the original author
- Le Théâtre de l’Est Parisien is the main theater company
- Guy Rétoré has multiple roles: he’s both director (“mise en scène”) and seems to have written the text
- It’s a French text, obviously, and the language is French
- The date range “12 juillet” gives us the first performance date
- The “Théâtre de l’Est Parisien” is clearly stated as the main company
- The fact it’s “de Sean O’Casey” tells us both the author and his work type
<|thinking_end|>
<|strategy_start|>
1. Title extraction: “Coquin de Coq” maps directly to P1705 (title)
2. Author identification: “de Sean O’Casey” indicates P50 (author)
3. Production details:
- “Mise en scène de Guy Rétoré” gives us:
* P57 (director) = Guy Rétoré
* P1705 (object has role) = “Mise en scène”
- “Le Théâtre de l’Est Parisien” maps to P272 (production company)
- French language maps to P407 (native label)
- First date “12 juillet” maps to P1191 (first performance date)
4. Creative roles:
- Guy Rétoré’s position as “Mise en scène” indicates P57 (director)
- His “texte français” suggests native language (P1705)
5. Format and context:
- Format indicates P31 (instance of) = “Works”
- Being a French text suggests P407 (native label) = “Coquin de Coq”
- Festival details map to P1191 (date of first performance)
- Source text maps to P248 (stated in)
<|strategy_end|>
<|data_start|>
<|subject|>Coquin de Coq
<|property|>title<|object|>{‘text’: ‘Coquin de Coq’, ‘language’: ‘fr”}
<|property|>native label<|object|>{‘text’: ‘Coquin de Coq’, ‘language’: ‘fr”}
<|property|>instance of<|object|>Works
<|property|>director<|object|>Guy Rétoré
<|property|>object has role<|object|>Mise en scène
<|property|>author<|object|>Sean O’Casey
<|property|>production company<|object|>Le Théâtre de l’Est Parisien
<|property|>date of first performance<|object|>12 juillet 1975
<|property|>stated in<|object|>Coquin de Coq show programme
<|data_end|>
\end{Verbatim}

\end{document}